\documentclass[sigconf]{acmart}

\usepackage{fontawesome}
\usepackage{amsmath,amsfonts,amscd}

\usepackage{amssymb}
\usepackage{fancyhdr}
\usepackage{mathrsfs}
\usepackage{graphicx}
\usepackage{textcomp}
\usepackage{subcaption}
\usepackage{multirow}
\usepackage{amsthm}
\usepackage[T1]{fontenc}
\usepackage{enumitem}
\usepackage{algorithm,algorithmicx,algpseudocode}
\usepackage{listings}
\usepackage{pifont}
\usepackage{booktabs}
\usepackage{threeparttable}
\newcommand{\macsection}[1]{\noindent\textbf{#1}~~~~}
\AtBeginDocument{%
  \providecommand\BibTeX{{%
    \normalfont B\kern-0.5em{\scshape i\kern-0.25em b}\kern-0.8em\TeX}}}

\acmConference[SC '21]{Supercomputing '21: The International Conference for High Performance Computing, Networking, Storage, and Analysis}{Nov 14–19, 2021}{St. Louis, MO, USA}
\acmYear{2021}
\acmDOI{10.1145/3458817.3476145}
\acmISBN{978-1-4503-8442-1/21/11}

\setcopyright{none}
\begin{document}

\title{Chimera: Efficiently Training Large-Scale Neural Networks with Bidirectional Pipelines}

\author{Shigang Li}
\email{shigangli.cs@gmail.com}
\affiliation{%
  \institution{Department of Computer Science, ETH Zurich}
  \country{Switzerland}
}

\author{Torsten Hoefler}
\email{htor@inf.ethz.ch}
\affiliation{%
  \institution{Department of Computer Science, ETH Zurich}
  \country{Switzerland}
}

\renewcommand{\shortauthors}{Shigang Li and Torsten Hoefler}

\begin{abstract}

Training large deep learning models at scale is very challenging. This paper proposes \textit{Chimera}, a novel pipeline parallelism scheme which combines bidirectional pipelines for efficiently training large-scale models. Chimera is a synchronous approach and therefore no loss of accuracy, which is more convergence-friendly than asynchronous approaches. Compared with the latest synchronous pipeline approach, Chimera reduces the number of bubbles by up to 50\%; benefiting from the sophisticated scheduling of bidirectional pipelines, Chimera has a more balanced activation memory consumption. Evaluations are conducted on Transformer based language models. For a GPT-2 model with 1.3 billion parameters running on 2,048 GPU nodes of the Piz Daint supercomputer, Chimera improves the training throughput by 1.16x-2.34x over the state-of-the-art synchronous and asynchronous pipeline approaches.

\end{abstract}




\maketitle

\section{Introduction}
\label{sec:intro}

Deep learning is continuing to deliver groundbreaking results on the path towards human-level intelligence. 
This path is characterized by growing model size, in just six years, the compute requirements for model training grew by 300,000 times~\cite{openai}. Transformers~\cite{vaswani2017attention} are a typical representative in this trend. As the model size grows, Transformer based models have proven their success in in the field of natural language processing~\cite{vaswani2017attention, devlin2018bert, radford2019language, radford2019language}.  Recent work~\cite{chen2020generative,dosovitskiy2020image,chen2020pre,carion2020end} shows that Transformers also achieve promising results in computer vision tasks, i.e., on par or better than other types of models such as
convolutional~\cite{krizhevsky2012imagenet} and recurrent~\cite{hochreiter1997long} networks. 
These growing models must be trained on distributed accelerator supercomputers.
Even today's models are too big to be stored on a single accelerator---for example, GPT-3's 175 billion parameters~\cite{brown2020language} require 350 GiB main memory if stored with 16 bits precision. Switch transformers~\cite{fedus2021switch} have in their largest configuration 1.6 trillion parameters, a 6.4 TiB storage requirements. 
Furthermore, the necessary memory for activations, gradients, and optimizer state during training at least triples these memory requirements.

Thus, full model replicas at each accelerator, as used in simple data parallelization schemes, are not sufficient. 
Large models must instead be distributed among many, often hundreds of accelerators to just fit into main memory.
Deep neural networks consist of a layered architecture and can thus be distributed in two ways: (1) the operators of a layer can be split across multiple accelerators in a method called \emph{operator parallelism} or (2) the model could be distributed layer by layer in a method called \emph{pipeline parallelism}~\cite{ben2019demystifying}. 
Operators of a typical fully-connected layer have computational characteristics similar to matrix multiplication, and splitting such a layer requires a communication volume of $\mathcal{O}(n^2/\sqrt{P})$~\cite{irony2004communication, kwasmmm} for an $n\times n$ matrix. By exploiting the inherent structure of Transformer based language models~\cite{shoeybi2019megatron}, operator parallelism requires two \textit{allreduce}~\cite{thakur2005optimization,li2013numa} operations on the output activations for each basic Transformer layer. Using a layer-wise model partition, pipeline parallelism on the other hand only requires point-to-point communication to transfer the output activations between pipeline stages, with each stage containing a group of consecutive layers. Therefore, pipeline parallelism commonly has a lower communication cost than operator parallelism. However, pipeline parallelism suffers from bubbles or weight staleness (see Section~\ref{background}), which are the problems this work aims to solve.
Overall, operator parallelism and pipeline parallelism are orthogonal and complementary to each other for distributing large deep learning models.

Yet, pipeline parallelism is not trivial: 
The backpropagation algorithm needs to ``remember'' the output activations computed during the forward pass as inputs to the backward pass (cf.~Figure~\ref{pipelinescompare}).
This creates wildly different memory requirements for each accelerator in the pipeline, even though each accelerator has an identical compute load.
Specifically, for some recently proposed pipeline approaches such as DAPPLE~\cite{fan2021dapple}, PipeDream~\cite{narayanan2019pipedream}, and PipeDream-2BW~\cite{narayanan2020memory}, the first accelerator of a pipeline of depth $D$ has to store $D$ such activations while the last accelerator requires memory for one. 
This does not only lead to lower memory utilization in the later pipeline stages (and only 50\% overall), it also leads to reduced performance because the micro-batch size has to be chosen to fit the first accelerator in the pipeline.
This imbalance can be alleviated by restricting the number of micro-batches that are simultaneously allowed in the pipeline.
However, this introduces bubbles and limits the overall system utilization.

Both micro-batch size and pipeline utilization are most important for the computational efficiency: larger micro-batches improve performance due to better re-use in the matrix-multiply-like operations and less pipeline bubbles (stalls) utilize the existing accelerators better.
The computational efficiency relates directly to the cost and time for of training a model.
We propose a new pipelining scheme, called \emph{Chimera}, that runs \textbf{fully-packed bidirectional pipelines} through the same set of accelerators. Chimera enables
\begin{itemize}
 \item to keep the overall training synchronous without relying on stale weights, 
 \item a higher pipeline utilization (less bubbles) than existing approaches and thus higher performance, 
 \item the same peak activation memory consumption as the state-of-the-art methods, with an extra benefit of more balanced memory consumption, and
 \item easy configurability to various pipelined deep neural networks as well as system architectures guided by an accurate performance model.
\end{itemize}

\begin{figure}[ht]
\centering\includegraphics[width=0.92\linewidth]{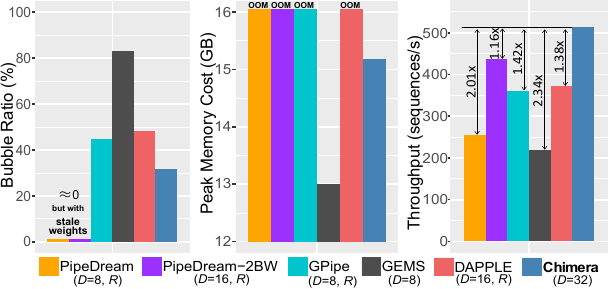}

\caption{\label{fig:poster} GPT-2 on 2,048 GPU nodes for mini-batch size=2,048. We show the bubble ratio (pipeline overhead), memory cost (OOM requires activation recomputation, denoted by $R$), and the best throughput ($D$ is the corresponding number of pipeline stages) for each approach. Details will be discussed in Sections~\ref{background} and~\ref{sec:parallelscala}.}

\end{figure}

For example, GPT-3 required 314 Zettaflop (mixed fp16/fp32) to train~\cite{brown2020language}, which would take a single A100 GPU more than 100 years.
The estimated cost to train GPT-3 varies between \$4.6m-\$12m.
We show that Chimera enables end-to-end performance improvements between 1.38x-2.34x per iteration for the synchronous training regime for a comparable GPT-2 model on 2,048 GPU nodes of the Piz Daint supercomputer, as shown in Figure~\ref{fig:poster}.
This enables savings of more than \$1.2m to \$5m when training very large models on practical systems.

\section{Background and Related Work}
\label{background}

Mini-batch stochastic gradient descent (SGD)~\cite{bottou2018optimization} is the mainstream method to train deep neural networks. Let $b$ be the mini-batch size, $w_t$ the neural network weights at step $t$, $(x_i, y_i)$ a sample in the mini-batch, and $\ell$ a loss function. During training, it computes the loss in the forward pass for each sample as $\ell(w_t, x_i, y_i)$, and then a stochastic gradient in the backward pass as
\[ g_t = \frac{1}{b} \sum_{i = 0}^b \nabla \ell(w_t, x_i, y_i). \] The model is trained in iterations such that $w_{t+1} = w_t - \eta_t g_t$. In more general terms, first-order stochastic gradient update rules can take different forms (e.g., by adding a momentum term), which is represented as $w_{t+1}=w_t + U\left(g_t,w_{(0,\dots,t)},t\right)$.

To scale up the training process to parallel machines, \textit{data parallelism}~\cite{sergeev2018horovod, you2018imagenet, goyal2017accurate, li2020taming} is the common method, in which the mini-batch is partitioned among $P$ workers and each worker maintains a copy of the entire model. Gradient synchronization across the workers is often implemented using an \textit{allreduce}. However, recent deep learning models~\cite{devlin2018bert, real2019regularized, radford2019language, brown2020language} scale rapidly from millions to billions of parameters. Pure data parallelism may not work for these large models since it either suffers from low efficiency caused by synchronizing gradients of the entire model across the workers or the model is simply too large to fit in a device.

\textit{Operator parallelism}
is a solution to train large models by partitioning the operators of a layer among multiple workers, but it may suffer from high communication volume as discussed in Section~\ref{sec:intro}. Hybrid approaches~\cite{krizhevsky2014one, jia2019beyond}, which combine operator parallelism with data parallelism, suffer from the similar problem to the pure operator parallelism. 

To reduce the communication volume of operator parallelism, \textit{pipeline parallelism} is intensively studied~\cite{narayanan2019pipedream, huang2019gpipe, jain2020gems, narayanan2020memory, fan2021dapple, wang2020fpdeep, gaunt2017ampnet, yang2019pipemare}. The key idea is to partition the model in a layer-wise way and treat each worker (and the layers on it) as a pipeline stage. The mini-batch is partitioned into multiple micro-batches, that are pipelined across the stages to increase the resources utilization. Recent work~\cite{fan2021dapple,jain2020gems,narayanan2020memory} also shows improved performance when combining pipeline parallelism with data parallelism. However, to efficiently pipeline deep neural network training is challenging because of (1) a training step contains one one forward pass followed by one backward pass, (2) the gradient computation in the backward pass rely on the intermediate results of the forward pass, and (3) to achieve good convergence accuracy the mini-batch size is usually not very large.

\begin{figure*}[ht!]
\centering\includegraphics[width=0.82\linewidth]{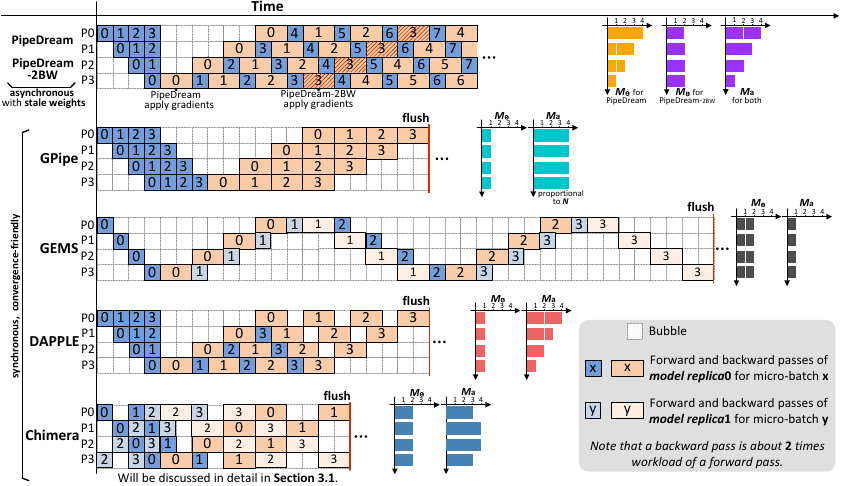}
\caption{\label{pipelinescompare} Pipeline parallelism schemes, with four pipeline stages ($D$=4) and four micro-batches ($N$=4) within a training iteration, except that PipeDream updates the model after each backward pass on a micro-batch.}
\end{figure*}

Table~\ref{tab:symbols} summarizes the symbols frequently used in the this paper. Next, we analyze the pros and cons of the state-of-the-art pipeline approaches when handling the challenges above from the aspects listed in Table~\ref{tab:pipeline-schemes}, and then show how our approach achieves the best balance among all aspects. To better understand the analysis in Table~\ref{tab:pipeline-schemes}, Figure~\ref{pipelinescompare} presents an example for each approach.

\begin{table}[h!]
  \caption{The list of symbols frequently used in the paper.}
  \label{tab:symbols}
  \centering
  \renewcommand{\arraystretch}{1.2}
  \begin{threeparttable}
  \begin{tabular}{p{0.5cm}p{7cm}}
    \toprule
    $D$ &  The number of pipeline stages (\textit{depth}) \\
    $W$ &  The number of replicated pipelines (\textit{width}) for data \ \ \ \ \ \ \ \ \ parallelism~\tnote{1} \\
    $P$ &  The number of workers ($= W * D$) \\
    $B$ &  Micro-batch size \\
    $N$ &  The number of micro-batches executed by each worker within a training iteration \\
    $\hat{B}$ &  Mini-batch size ($= B * N * W$) \\
    $M_{\theta}$ &  Memory consumption for the weights of one stage \\
    $M_{a}$ &  Memory consumption for the activations of one stage \\
    \bottomrule
  \end{tabular}
  \begin{tablenotes}
       \footnotesize
       \item[1] This paper considers the cases where all pipeline stages have balanced workload, and therefore are equally replicated to combine with data parallelism. 
    \end{tablenotes}
  \end{threeparttable}
\end{table}

\begin{table*}[ht!]
  \caption{Comparison between different pipeline schemes.}
  \label{tab:pipeline-schemes}
  \centering
  \renewcommand{\arraystretch}{1.2}
  \begin{threeparttable}
  \begin{tabular}{lclrl}
    \toprule
    Pipeline Schemes &  Bubble Ratio & Weights Memory & Activations Memory \ \ \ \  & \  Convergence Friendly \\
    \midrule
    
    PipeDream~\cite{narayanan2019pipedream} & $\approx 0$ \ \  \faThumbsOUp \faThumbsOUp & $[M_{\theta},\  D*M_{\theta}]~\tnote{1}$\ \  \faThumbsDown & $[M_{a},\  D*M_{a}]~\tnote{1}$ \ \ \ \ \  \faThumbsOUp & \multirow{2}{*}{ \ \ \ \ \ Asynchronous \faThumbsDown } \\
    
    PipeDream-2BW~\cite{narayanan2020memory} & $\approx 0$ \ \  \faThumbsOUp \faThumbsOUp & \ \ \ \ \ \ \ \ \ $2M_{\theta}$ \ \ \faThumbsOUp & $[M_{a},\  D*M_{a}]~\tnote{1}$ \ \ \ \ \  \faThumbsOUp &  \\   
    \midrule
    GPipe~\cite{huang2019gpipe} &  $(D-1)/(N+D-1)$ \ \ \ \ \faThumbsDown &  \ \ \ \ \ \ \ \ \ \  $M_{\theta}$ \ \ \ \faThumbsOUp & $N*M_{a}$  \ \ \ \ \ \ \ \faThumbsDown \faThumbsDown & \multirow{4}{*}{ \ \ \ \ \ Synchronous \ \   \faThumbsOUp} \\
    
    GEMS~\cite{jain2020gems} & $ \approx (D-1)/(D+\frac{1}{2})$ \ \  \faThumbsDown \faThumbsDown & \ \ \ \ \ \ \ \ \ $2M_{\theta}$ \ \ \faThumbsOUp & $M_{a}$  \ \ \ \ \ \ \ \ \ \ \ \faThumbsOUp \faThumbsOUp & \\
    
    DAPPLE~\cite{fan2021dapple} & $(D-1)/(N+D-1)$ \ \ \ \ \faThumbsDown &  \ \ \ \ \ \ \ \ \ \ $M_{\theta}$ \ \ \  \faThumbsOUp & $[M_{a},\ D*M_{a}]~\tnote{1}$ \ \ \ \ \  \faThumbsOUp & \\
    
    \textbf{Chimera (this work)} &  $(D-2)/(2N+D-2)$ \ \ \faThumbsOUp &  \ \ \ \ \ \ \ \ \ $2M_{\theta}$ \ \ \faThumbsOUp & $[(D/2+1)M_{a}, D*M_{a}]~\tnote{1}$ \  \faThumbsOUp + & \\

    \bottomrule
  \end{tabular}
  \begin{tablenotes}
       \footnotesize
       \item[1] Intervals for the values across the workers.
    \end{tablenotes}
  \end{threeparttable}
\end{table*}

\macsection{Bubbles in the pipeline.} For better convergence quality, synchronous approaches synchronize the gradients and flush the pipeline at the end of each training iteration, as shown in Figure~\ref{pipelinescompare}. Therefore, synchronous approaches lead to pipeline bubbles.  To utilize the pipeline, GPipe~\cite{huang2019gpipe} injects $N$ micro-batches into the pipeline concurrently; DAPPLE~\cite{fan2021dapple} uses the One-Forward-One-Backward (1F1B~\cite{ narayanan2019pipedream,narayanan2020memory}) schedule with periodic flushes. Both GPipe and DAPPLE incur 2($D$-1) bubbles (i.e., $D$-1 bubbles in the forward passes and $D$-1 bubbles in the backward passes). In contrast, Chimera (this work) incurs $D$-2 bubbles (i.e., $D/2$-1 bubbles in the forward passes and $D/2$-1 bubbles in the backward passes), which is about 50\% reduction compared with DAPPLE and GPipe. We define the \textit{bubble ratio} as the bubble overhead divided by the overall runtime of the pipeline. In practice, the workload of a backward pass is about two times of a forward pass, which leads to $D/2$-1 bubbles in the middle of the pipeline for Chimera (bubble ratio = ($D$-2)/(3$N$/2+$D$-2)), as shown in Figure~\ref{pipelinescompare}. We will discuss how to remove the middle bubbles in Section~\ref{sec:largerbatch} (bubble ratio = ($D$-2)/(2$N$+$D$-2)).

Table~\ref{tab:pipeline-schemes} presents the bubble ratio for all approaches with the consideration of typical workloads of forward and backward passes.
Although the bubble ratio of GPipe and DAPPLE decreases as $N$ (the number of micro-batches executed by each worker within a training iteration) increases, a large enough $N$ ($N$>=4$D$ as suggested in ~\cite{huang2019gpipe}) usually cannot be obtained without hurting the efficiency for the following three reasons: (1) There is usually an empirical maximum $\hat{B}$ (mini-batch size) for a model, exceeding which would compromise model convergence~\cite{you2018imagenet,you2019largebert,you2019largelstm,ben2019demystifying,cheng2021dataset}. (2) An increase of $N$ implies an decrease of $B$ (micro-batch size) for a given $\hat{B}$. However, modern accelerators require a large enough $B$ to achieve high computational efficiency. (3) Scaling to large-scale machines by combining with data parallelism (which has proven to be an effective way~\cite{narayanan2020memory, fan2021dapple}) would decrease $N$ for a given $\hat{B}$.

\begin{figure*}[!ht]
\centering\includegraphics[width=0.85\linewidth]{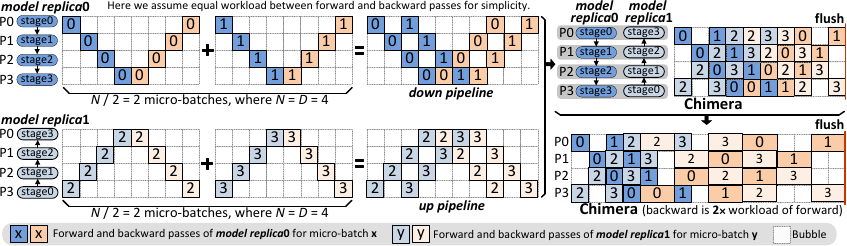}
\caption{\label{chimeraschedule} Model replicas and bidirectional pipelines scheduling of Chimera.}
\end{figure*}

GEMS~\cite{jain2020gems} is a memory-efficient pipeline approach which schedules micro-batches among two model replicas. Since GEMS is mainly designed for small $\hat{B}$ and has at most two active micro-batches at the same time, its bubble ratio is much higher than the other approaches and cannot be alleviated by a larger $N$. Asynchronous approaches (such as PipeDream~\cite{narayanan2019pipedream} and PipeDream-2BW~\cite{narayanan2020memory}) do not have periodic pipeline flushes, so they do not have bubble problem but with stale weights.

\macsection{Memory consumption.} Memory overhead mainly comes from two aspects: the weight parameters and the activations (the intermediate results of the forward pass required in the backward pass to compute gradients). For GPipe and DAPPLE, each worker maintains the weights of one pipeline stage. For GEMS and Chimera (with the default setting), each worker maintains the weights of two pipeline stages since there are two pipelines in two directions. PipeDream updates the model after each backward pass on a micro-batch ($N$=1); therefore, to ensure weight version consistency between forward and backward passes, it requires to stash up to $D$ versions of weights on a worker, which has the same memory cost as pure data parallelism. By using gradient accumulation ($N$>=$D$), PipeDream-2BW reduces the number of weight versions to be stashed to 2.

GEMS injects only one micro-batch at the beginning of the pipeline, and thus the activations of the forward pass on one micro-batch are stored. However, this leads to low pipeline efficiency as discussed previously. Since GPipe injects $N$ micro-batches ($N>=D$ to fully utilize the pipeline) into the pipeline concurrently, the activitions memory consumption is proportional to $N$, which does not scale well to large mini-batches. Using the classic (or a variant of) 1F1B~\cite{ narayanan2019pipedream,narayanan2020memory} schedule, PipeDream, PipeDream-2BW, DAPPLE, and Chimera inject up to $D$ micro-batches at the beginning of the pipeline, which scale well to large mini-batches. By counting the number of injected micro-batches on each worker of Chimera in Figure~\ref{pipelinescompare}, we can observe that Chimera has an extra benefit of a more balanced activations memory consumption among the workers (see Table~\ref{tab:pipeline-schemes} for the general analysis) than PipeDream, PipeDream-2BW, and DAPPLE, and therefore a better memory resources utilization.  
Note that the activations memory consumption can be reduced using the technique of \textit{activation recomputation}~\cite{chen2016training}, but this is at a cost of about 1/3 more computation overhead~\cite{fan2021dapple,narayanan2020memory}.

ZeRO~\cite{rajbhandari2020zero, ren2021zero} removes the
memory redundancies by partitioning
the three model states (i.e., optimizer states, gradients, and
parameters) across data-parallel processes, with a modest increasement to the communication volume. Note that our pipeline approach is orthogonal to ZeRO. To further reduce the memory consumption of our pipeline approach is an interesting future work.

\macsection{Convergence friendliness.} By periodic pipeline flushes, synchronous approaches ensure that the same version of weights is used across all stages and all micro-batches in a training iteration, without introducing staleness. From the algorithmic perspective, synchronous approaches are equivalent to the standard and well-proved mini-batch SGD, and therefore, guarantee convergence.

To remove pipeline flushes, asynchronous approaches either aggressively lead to weight versions mismatch between forward and backward passes (such as AMPNet~\cite{gaunt2017ampnet} and PipeMare~\cite{yang2019pipemare}), or conservatively introduce staleness to the weights while ensuring weight versions consistency (such as PipeDream-2BW and PipeDream). Although they empirically show promising convergence results, the generality is lack of proof. More recent work~\cite{asyncring,assran2018stochastic,nadiradze2019swarmsgd,tang2020,li2020breaking} observes that asynchronous training algorithms may result in lower convergence performance.

For the model accuracy, all the synchronous pipeline approaches (such as Chimera, DAPPLE, GPipe and GEMS) are guaranteed to achieve the same accuracy as the standard mini-batch SGD algorithm. For the asynchronous approaches (such as PipeDream-2BW and PipeDream), it is not safe to achieve the ideal accuracy as the standard algorithm because of the introduced weight staleness, and the convergence quality may exhibit variance on different neural networks and tasks.

Overall, Chimera achieves the best balance of all aspects, as presented in Table~\ref{tab:pipeline-schemes}. We will discuss the implementation details of Chimera in the following section.

\section{The scheme of Chimera}

\subsection{Bidirectional Pipelines}
\label{sec:bipipes}

We consider large-scale models with repetitive structures (i.e., the same block repeated
multiple times), such as Bert~\cite{devlin2018bert} and GPT-2/3~\cite{radford2019language, brown2020language}, which can be partitioned into balanced stages with an equal number of blocks. The feature of repetitive structures is also exploited in PipeDream-2BW~\cite{narayanan2020memory}. How to generally partition any model into stages with efficiency is not the topic of this paper and has been well studied in recent work~\cite{narayanan2019pipedream,fan2021dapple}.

The key idea of Chimera is to combine two pipelines in different directions (we call them \textit{down} and \textit{up} pipelines, respectively) together. Figure~\ref{chimeraschedule} shows an example with four pipeline stages (i.e., $D$=4). Here we assume there are $D$ micro-batches executed by each worker within a training iteration, namely $N$=$D$, which is the minimum to keep all the stages active. How to scale to more than $D$ micro-batches (i.e., for $N$>$D$) will be discussed in Section~\ref{sec:largerbatch}. In the \textit{down} pipeline, \textit{stage}0$\sim$\textit{stage}3 are mapped to \textit{P}0$\sim$\textit{P}3 linearly, while in the \textit{up} pipeline the stages are mapped in a completely opposite order. The $N$ (assuming an even number) micro-batches are equally partitioned among the two pipelines. Each pipeline schedules $N/2$ micro-batches using 1F1B~\cite{narayanan2019pipedream} strategy, as shown in the left part of Figure~\ref{chimeraschedule}. Then, by merging these two pipelines together, we obtain the pipeline schedule of Chimera (upper right of Figure~\ref{chimeraschedule}). Given an even number of stages $D$ (which can be easily satisfied in practice), it is guaranteed that there is no conflict (i.e., there is at most one micro-batch occupies the same time slot on each worker) during merging. We can see that the number of bubbles is reduced to $D$/2-1 in the forward and backward passes, respectively. By considering the uneven workloads between forward and backward passes, we get a more practical schedule of Chimera (bottom right of Figure~\ref{chimeraschedule}).

For the models which have to use a small $\hat{B}$ to guarantee convergence, there maybe less than $D$ micro-batches in a training iteration (i.e., $N$<$D$). Chimera also supports the cases of $N$<$D$ by simply partitioning the $N$ micro-batches among the two pipelines as evenly as possible, with an extreme case that $N$=1 where only one micro-batch runs on a single pipeline.

Note that Chimera can be generalized to combine more than two pipelines (will be discussed in Section~\ref{sec:morepipes}), which further reduces the bubbles and balances the activations memory consumption, but at a cost of higher communication overhead and weights memory consumption. 

\begin{figure}[ht]
\centering\includegraphics[width=0.88\linewidth]{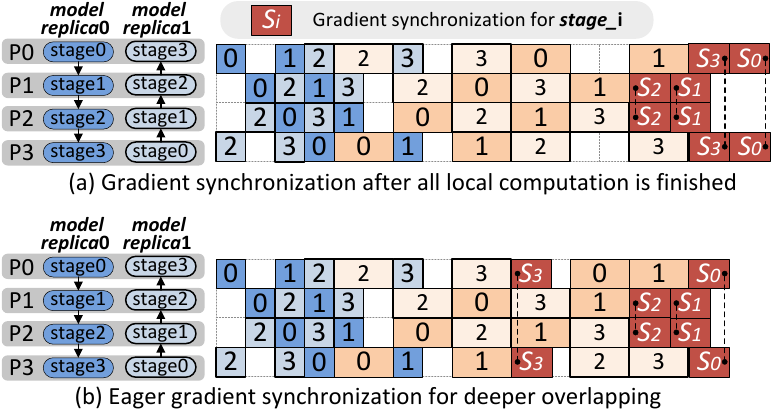}
\caption{\label{commopt} Hide the gradient synchronization overhead by overlapping.}
\end{figure}
\vspace{-1em}

\subsection{Communication Scheme}
\label{sec:commscheme}

Chimera uses \textit{p2p} (point-to-point) communication to transfer the intermediate activations and gradients (with respect to the inputs) between pipeline stages in the forward pass and the backward pass, respectively. Since Chimera combines bidirectional pipelines together, collective communication (i.e., \textit{allreduce}) is used to synchronize the weight gradients across stage replicas before the next training iteration. Figure~\ref{commopt}(a) presents a simple way for gradient synchronization, namely synchronizing the gradients for each stage maintained by the workers after all the local computation of the current iteration is finished. Note that the gradient synchronization for the middle stages is partially overlapped by the computation on the beginning and the end stages. 

For a deeper communication overlap, we launch \textit{allreduce} eagerly by utilizing the bubbles in the pipeline. Taking P0 and P3 in Figure~\ref{commopt}(b) as an example, after these two workers finish the backward passes on \textit{micro-batch} 3 and \textit{micro-batch} 1, respectively, the calculation for the weight gradients of \textit{stage}3 has been finished; therefore, P0 and P3 can launch an asynchronous \textit{allreduce} using nonblocking collectives~\cite{hoefler2007case,hoefler-sc07} to synchronize the gradients of \textit{stage}3 as soon as they are finished, and a \textit{wait} operation is called after all the local computation to make sure the \textit{allreduce} is finished. In this way, the gradient synchronization for \textit{stage}3 is overlapped by the bubbles and the following computation. However, unlike P0 and P3, we choose not to conduct eager gradient synchronization for \textit{stage}2 (a middle stage) on P1 and P2, since there is no bubble from the completion of \textit{stage}2's gradients to the end of local computation. Although the asynchronous communication may proceed while the computation happens, it may cause additional overheads (initialization, threading etc.~\cite{hoefler-ib-threads}), which could extend the critical path of the pipeline and jeopardize the overall performance. Performance modelling of the communication scheme will be presented in Section~\ref{sec:perfmodel}.

\begin{figure}[ht]
\centering\includegraphics[width=0.82\linewidth]{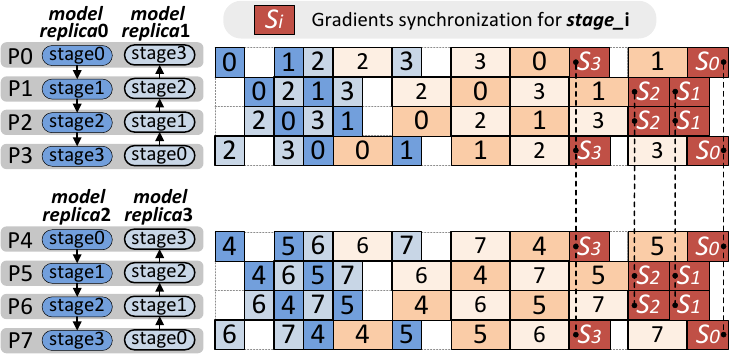}
\caption{\label{hybridp} Combine pipeline parallelism with data parallelism in Chimera ($W=2, D=4$).}
\end{figure}
\vspace{-1em}

\subsection{Hybrid of Pipeline and Data Parallelism}

Chimera supports a hybrid of pipeline and data parallelism. The bidirectional pipelines of Chimera are replicated $W$ times to scale to $W\cdot D$ workers. Since we consider the large models which can be easily partitioned into balanced stages, all $D$ stages are equally replicated $W$ times in hybrid parallelism. When scaling to the parallel machines equipped with high performance interconnected networks (such as Infiniband~\cite{shanley2003infiniband}, Cray Aries~\cite{alverson2012cray} or Slingshot~\cite{slingshot}, and NVLink~\cite{foley2017ultra}), hybrid parallelism usually achieves better performance than the pure pipeline parallelism~\cite{fan2021dapple,narayanan2020memory}. This is because pure pipeline parallelism has $W\cdot D$ stages in the pipeline, while hybrid parallelism has $D$ stages ($W$ times less) which helps to reduce the \textit{p2p} communication overhead between stages and increase the computation workload of each stage. Although hybrid parallelism leads to gradient synchronization between stage replicas, the overhead of it can be alleviated by the aforementioned high performance interconnected networks. However, as $W$ increases ($D$ decreases), pipeline stages become coarser, until at some point the increased gradient synchronization overhead cannot be amortized by the reduced \textit{p2p} communication overhead. Therefore, it is important to find the sweet spot to achieve the best performance. 

Figure~\ref{hybridp} presents an example with $W$=2 and $D$=4. Note that after combining with data parallelism, the size of local gradients to be synchronized does not change, but the number of processes participating in the gradient synchronization increases by $W$ times. Also, we use the same communication scheme as discussed in Section~\ref{sec:commscheme} to overlap the gradient synchronization overhead in hybrid parallelism. In the next section we will discuss how to find the best configuration of $W$ and $D$ based on performance modelling.

\subsection{Configuration Selection Based on Performance Modelling}
\label{sec:perfmodel}

Given the mini-batch size $\hat{B}$ and the number of workers $P$, the configuration of $B$, $W$, and $D$ largely affects the training throughput.

Larger micro-batch size ($B$) usually improves the computational efficiency of the accelerators. Since Chimera greatly alleviates the bubble problem, it greedily chooses to use the maximum micro-batch size fitting in the device memory. Compared with the existing synchronous pipeline approaches which have to consider the trade-off between bubble and computational efficiency, the greedy strategy of Chimera significantly reduces the tuning space.

\begin{figure}[h!]
\centering\includegraphics[width=0.92\linewidth]{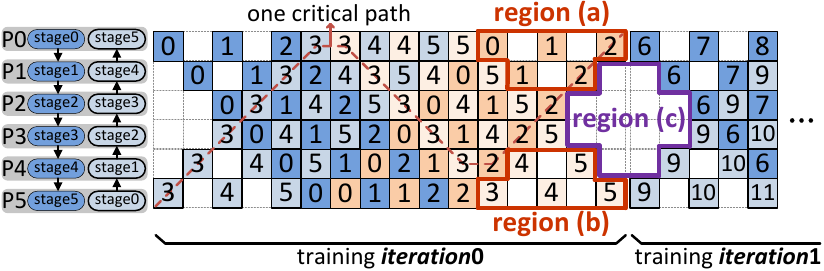}
\caption{\label{freeregion} The free regions to overlap gradient synchronization in Chimera, where $N=D=6$.}
\end{figure}

To select the best configuration of $W$ and $D$, we build a performance model to predict the runtime of a single training iteration (represented by $T$) for each available configuration. For the computation overhead, we measure the runtime of the forward pass on a single pipeline stage (represented by $F_{t}$) using micro benchmarks. The runtime of backward pass (represented by $B_{t}$) is modelled as two times of the forward pass if no activation recomputation is used, and three times otherwise. We define the \textit{critical path} as a series of pipeline stages with dependency that determine the overall computation overhead of a single training iteration. An example of critical path is shown in Figure~\ref{freeregion}. Let $C_{f}$ and $C_{b}$ denote the number of forward passes and backward passes on the critical path of the pipeline, respectively. For the example shown in Figure~\ref{freeregion}, $C_{f}$=6 and $C_{b}$=10. The total computation overhead is $F_{t}C_{f}$+$B_{t}C_{b}$.

To model the communication overhead, we assume bidirectional and direct point-to-point communication between the compute nodes, and use the classic Latency-Bandwidth ($\alpha - \beta$) cost model. The cost of sending a message of size $L$ is $\alpha + \beta L$, where both $\alpha$ (latency) and $\beta$ (the transfer time per word) can be measured using micro benchmarks. As discussed in Section~\ref{sec:commscheme}, Chimera has two types of communication: \textit{p2p} communication ($Comm_{p2p}$) between stages and \textit{allreduce} ($Comm_{allreduce}$) for gradient synchronization. $Comm_{p2p}$ can be simply modelled by the $\alpha - \beta$ cost model. The total \textit{p2p} communication cost is $(C_{f}+C_{b})Comm_{p2p}$. Note that $Comm_{p2p}$ can be partially overlapped by the intermediate bubbles if there are any, but to simplify the model we do not consider that effect.

For $Comm_{allreduce}$, we assume its implementation makes use of Rabenseifner’s algorithm~\cite{rabenseifner2004optimization,thakur2005optimization}, whose cost is $$Comm_{allreduce} = 2(log_{2}r)\alpha+2(r-1)\beta L/r$$

\noindent where $L$ is the size of gradients to be synchronized and $r$ is the number of stage replicas. Note that Rabenseifner’s algorithm reaches the lower bound on the bandwidth term for host-based \textit{allreduce}, and therefore works best for large models. We model the effect of communication overlapping (discussed in Section~\ref{sec:commscheme}) for $Comm_{allreduce}$. Figure~\ref{freeregion} shows an example of the free  regions (i.e., exceeding which will increase the total runtime) utilized in Chimera to overlap the gradient synchronization. Note that there are two stage replicas on each worker. \textit{Regions} (a) and (b) can be utilized to overlap the gradient synchronization for the first stage replica (the one with a larger stage ID), and \textit{region} (c) can be utilized to overlap the gradient synchronization for both stage replicas. Let $Comm_{unoverlapped}(i)$ represent the portion of $Comm_{allreduce}$ which can not be overlapped by the free regions on worker $i$, and then the max of $Comm_{unoverlapped}(i)$ among the $D$ workers contributes to the total runtime.

Overall, the runtime of a single training iteration is modelled as 
\begin{align}
T = (F_{t} + Comm_{p2p})C_{f} + (B_{t} + Comm_{p2p})C_{b} + \nonumber \\ \quad\quad\quad\quad max\{Comm_{unoverlapped}(i): i \in [0, D-1]\}.
\label{equ:perfmodel}
\end{align}
We use this model to select the best configuration of $W$ and $D$ (see Section~\ref{sec:perfmodelexperiment}).

\begin{figure}[ht]
\centering\includegraphics[width=0.96\linewidth]{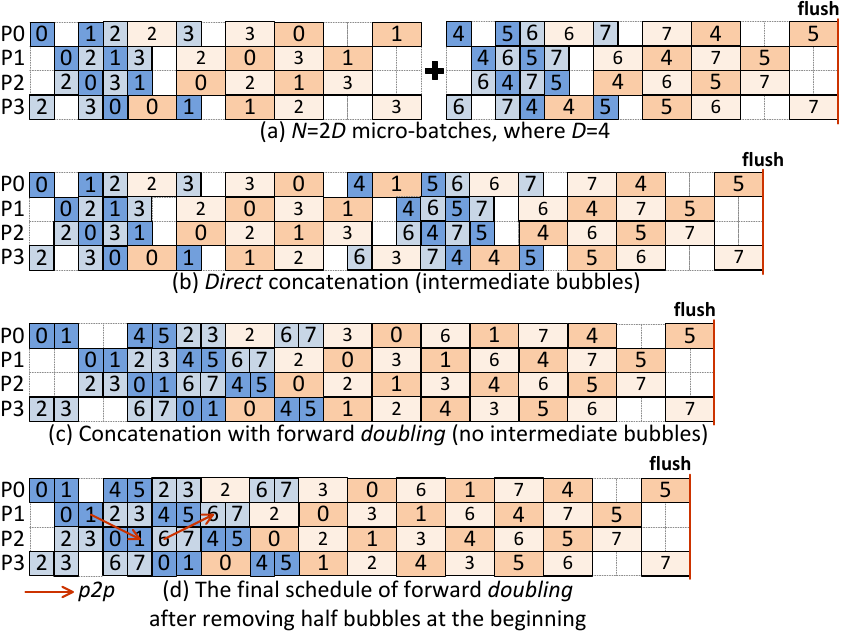}
\caption{\label{scalebatches} Scale to more than $D$ micro-batches within a training iteration (i.e., $N>D$) for Chimera.}
\end{figure}
\vspace{-1em}

\begin{figure*}[ht]
\centering\includegraphics[width=0.99\linewidth]{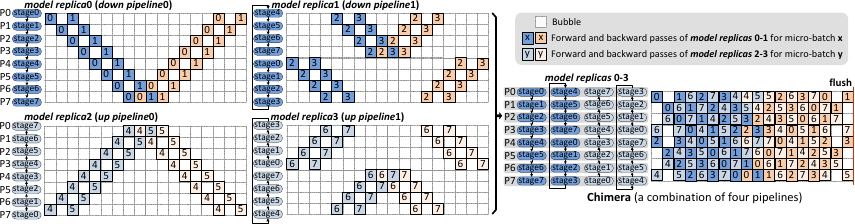}
\caption{\label{morereplicas} Chimera with a combination of four 8-stage pipelines.}
\end{figure*}

\subsection{Scale to More Micro-Batches}
\label{sec:largerbatch}

For a large $\hat{B}$, there may be more than $D$ micro-batches in a training iteration for each worker (i.e., $N$>$D$), especially when the compute resources are limited. To scale to a large $\hat{B}$, we first choose the maximum $B$ with $D$ micro-batches to saturate the device memory, and schedule these $D$ micro-batches using bidirectional pipelines as discussed previously. Then, we use the schedule of $D$ micro-batches as a basic scheduling unit, and scale to a large $\hat{B}$ by concatenating $K$ ($K$=$N/D$ and $N$=$\hat{B}/W/B$) basic units together. Figure~\ref{scalebatches}(a) presents an example with $N$=$2D$ micro-batches in a training iteration for each worker, which has two basic scheduling units (i.e., $K$=2). We propose three methods to concatenate multiple basic units. (1) \textit{Direct} concatenation, as shown in Figure~\ref{scalebatches}(b). The bubbles at the end of the first basic unit can be occupied by the forward passes at the beginning of the second basic unit. If the backward pass has the same workload as the forward pass, basic units can be concatenated seamlessly. However, backward pass has about two times workload of the forward pass, which results in intermediate bubbles.

To remove the intermediate bubbles of \textit{direct} concatenation, we propose (2) forward \textit{doubling} (shown in Figure~\ref{scalebatches}(d)) and (3) backward \textit{halving}, in which the key idea is to equalize the workloads of forward and backward passes. In forward \textit{doubling}, we increase the number of micro-batches for each forward pass to two, and then concatenate the two consecutive chunks of backward passes, each of which has only one micro-batch, as shown in Figure~\ref{scalebatches}(c). Then, we fine-tune the schedule to remove 50\% bubbles at the beginning of the pipeline, as shown in Figure~\ref{scalebatches}(d). Forward \textit{doubling} removes the intermediate bubbles, but it leads to two times activation memory consumption and therefore may exceed the device memory capacity. We resort to \textit{activation recomputation} to reduce memory overhead. Note that recomputation increases the workload of the backward pass, but the \textit{p2p} communication overhead in the forward passes is also doubled because of the outputs for two micro-batches; therefore, we still treat the forward pass (with two micro-batches) and the backward pass (with one micro-batch and recompute) have approximately equal workload. Forward \textit{doubling} prefers large models in which even $B$=1 exceeds the device memory capacity, since in such case activation recomputation must be used. For smaller models which has a larger $B$, we propose to use backward \textit{halving}, which uses the same schedule as forward \textit{doubling}, except that rather than executing two micro-batches in the forward pass but to halve the micro-batch size of the backward pass. Backward \textit{halving} does not increase the activation memory (thus no activation recomputation), but it may lower the computational efficiency because of using a sub-max $B$. To select the best of the three methods is not a priori, which we rely on empirical results. Note that both forward \textit{doubling} and backward \textit{halving} have total $D$-2 bubbles ($D$/2-1 in the forward passes and $D$/2-1 in the backward passes), as shown in Figure~\ref{scalebatches}(d), which is about a 50\% reduction compared with DAPPLE and GPipe. For $K$>2, we use the schedule of 2$D$ micro-batches as a basic scheduling unit (as shown in Figure~\ref{scalebatches}(c)) for forward \textit{doubling} and backward \textit{halving}, and concatenate $\lfloor k/2 \rfloor$ basic units and the residual $D$ micro-batches if $K$ is odd.

One more benefit for both forward \textit{doubling} and backward \textit{halving} is that they have more space to overlap \textit{p2p} communication (in the forward passes) than the classic 1F1B schedule~\cite{ narayanan2019pipedream,narayanan2020memory}. In Figure~\ref{scalebatches}(d), taking the forward pass on \textit{micro-batch} 1 as an example, the \textit{p2p} communication from P1 to P2 can be overlapped by the intermediate forward pass computation, while for 1F1B there may be not enough computation to overlap \textit{p2p} communication.

\subsection{Generalize to More than Two Pipelines}
\label{sec:morepipes}

So far we have only discussed the case that two pipelines (one \textit{down} and one \textit{up}) are combined together in Chimera. Yet, Chimera can be generalized to combine more than two pipelines for an even number of pipeline stages (i.e., $D$ is even). For $Q$=$D$/2, let $F$ denote the set of all the divisors of $Q$, including 1 and $Q$ itself. For any $f\in F$, we can generate a scheme for Chimera, which combines $f$ \textit{down} pipelines and $f$ \textit{up} pipelines together and each pipeline has $D$/2$f$ micro-batches scheduled by the 1F1B strategy. Figure~\ref{morereplicas} gives an example in which Chimera combines four pipelines with eight stages (i.e., $D$=8 and $f$=2). For the \textit{down pipeline} $i$ ($i \in [0, f-1]$), the $D$ stages are mapped to the $D$ workers in turn with the first stage (i.e., \textit{stage}0) being mapped to the worker $i*(D/f)$. For example, for the \textit{down pipeline}1 in Figure~\ref{morereplicas}, stages [0,1,2,3,4,5,6,7] are mapped to workers [4,5,6,7,0,1,2,3], respectively. For the $f$ \textit{up} pipelines, the $D$ stages are mapped to the $D$ workers in a completely reverse order of the corresponding \textit{down} pipeline. It can be easily demonstrated that the schedules of the $2f$ pipelines can be overlaid without conflict.

\begin{table}[h!]
  \caption{Chimera with 2$f$ pipelines.}
  \label{tab:chimeramorepipes}
  \centering
  \renewcommand{\arraystretch}{1.2}
  \begin{tabular}{lc}
    \toprule
    Model Replicas &  2$f$  \\
    Bubble Ratio &  $(D-2f)/(2fN+D-2f)$ \\
    Weights Memory &  $2f*M_{\theta}$ \\
    Activations Memory &  $[(D-D/2f+1)M_{a},\ \  D*M_{a}]$ \\
    \bottomrule
  \end{tabular}
\end{table}

For any $f\in F$, Chimera can scale to more micro-batches (i.e., $N$>$D$) using the methods discussed in Section~\ref{sec:largerbatch}. For a given $f$, Chimera incurs 2($D$/$f$/2-1) bubbles, but has to maintain $2f$ model replicas and synchronize the gradients of $2f$ stages on each worker. The larger the value of $f$, the less bubbles (and the more balanced activations memory consumption), but also the higher gradient synchronization overhead and weights memory consumption. When $f$=$Q$, Chimera degrades to pure data parallelism. Empirical results in Section~\ref{eval:morepips} show that $f$>1 rarely brings more performance benefit on the models used for evaluation. Thus, $f$=1 (i.e., a combination of two pipelines) is the default setting for Chimera in this paper, unless otherwise stated. We expect that $f$>1, whose features are summarized in Table~\ref{tab:chimeramorepipes}, would further improve the performance for future deep models with deeper pipeline.

\section{Experimental Evaluation}

We conduct our experiments mainly on the CSCS Piz Daint supercomputer. Each Cray XC50 compute node contains an Intel Xeon E5-2690 CPU, and one NVIDIA P100 GPU with 16 GB global memory. The compute nodes of Piz Daint are connected by Cray Aries interconnect network in a Dragonfly topology. 

We also evaluate the performance on a small cluster equipped with 32 NVIDIA V100 GPUs. The cluster has four GPU servers connected by Infiniband, and each server has eight V100 GPUs with NVLink. Each V100 GPU has 32 GB global memory.

\begin{figure}[ht]
\centering\includegraphics[width=0.95\linewidth]{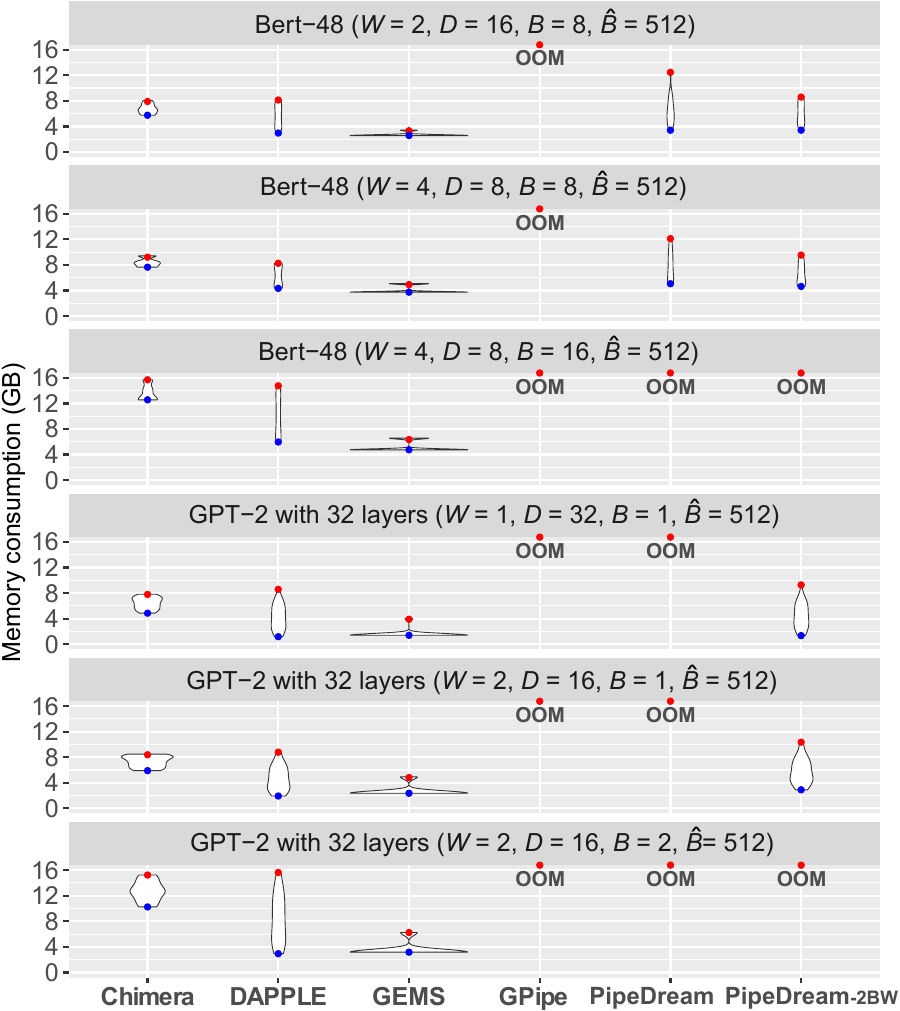}
\caption{\label{memCons} Memory consumption distribution among 32 GPU nodes of Piz Daint. The red and blue circles indicate the maximum and minimum memory consumption, respectively.}
\end{figure}

\begin{table}[h!]
  \caption{Neural networks used for evaluation.}
  \label{tab:networks}
  \centering
  \renewcommand{\arraystretch}{1.2}
  \begin{tabular}{lclc}
    \toprule
    Networks & Layers & Parameters & Mini-batch size \\
    \midrule
    Bert-48 & 48 & 669,790,012 & >=256 \\
    GPT-2 & 64 & 1,389,327,360 & >=512 \\
    \bottomrule
  \end{tabular}
\end{table}
\vspace{-1em}

We evaluate the performance of the schemes listed in Table~\ref{tab:pipeline-schemes}, which covers the state-of-the-art. For a fair comparison, all schemes are implemented in PyTorch~\cite{paszke2019pytorch} with GLOO~\cite{gloo} distributed backend for both the point-to-point (\textit{p2p}) communication between pipeline stages and gradient synchronization (\textit{allreduce}) across the stage replicas, and GPU devices are utilized for acceleration. Although NCCL~\cite{nccl} backend of PyTorch performs better for \textit{allreduce} across GPU nodes (with GPUDirect RDMA), it does not support \textit{p2p} communication. Using NCCL for gradient synchronization and GLOO for \textit{p2p} at the same time fails, which is also observed in PipeDream~\cite{narayanan2019pipedream}. We use the language models summarized in Table~\ref{tab:networks} for evaluation, and the max sequence length of Bert-48 and GPT-2 are set to 128 and 632 respectively, unless otherwise stated. The mini-batch size and sequence length we use are consistent with those in the machine learning community~\cite{devlin2018bert,radford2019language,you2019largebert,wolf2019huggingface}. Since Chimera is a synchronous approach without compromising convergence accuracy, we focus on the training throughput comparison.

\begin{figure*}[h!]
\centering\includegraphics[angle=90, width=0.82\linewidth]{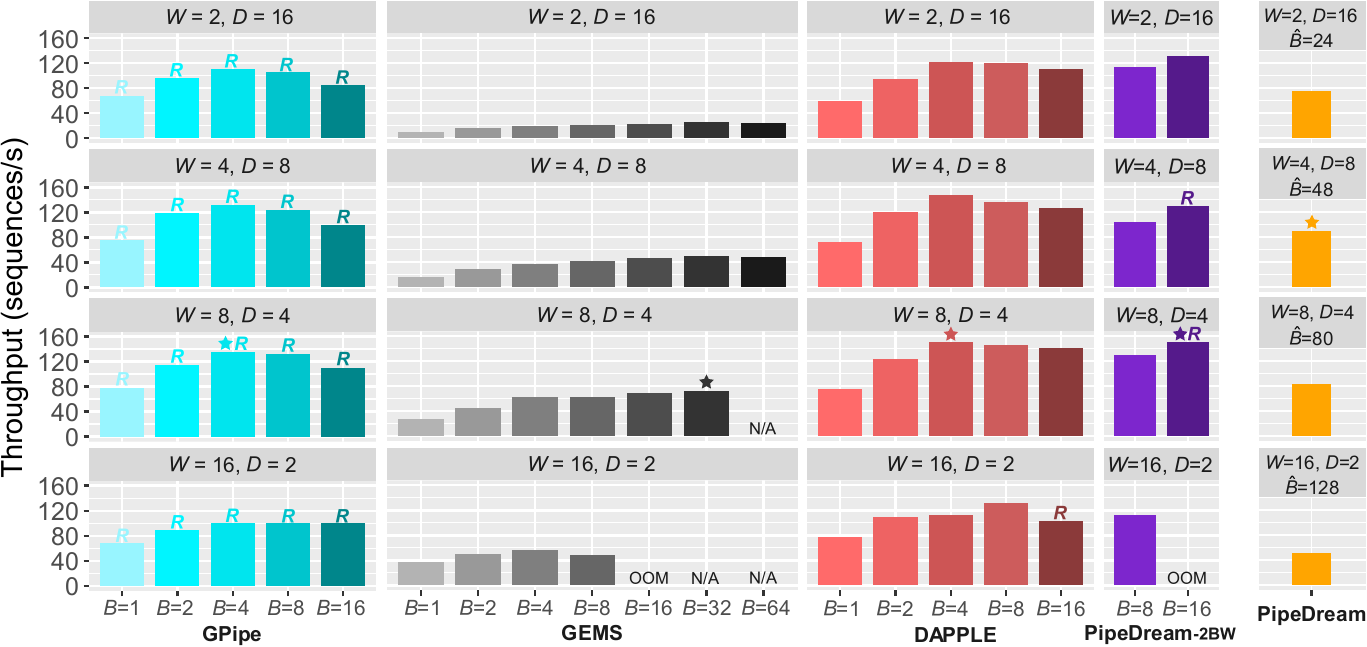}
\caption{\label{bertbaselinetuning} Performance tuning for the baselines for Bert-48 on 32 GPU nodes of Piz Daint. $\hat{B}$=512, except for PipeDream uses the maximum $\hat{B}$ fitting in the memory. $R$ denotes activation recomputation to avoid \textit{OOM}. $Star$ marks the best performance.}
\end{figure*}

\subsection{Memory Consumption}
\label{sec:memcons}

Figure~\ref{memCons} presents the memory consumption (including both activations and weights) distribution among 32 GPU nodes of Piz Daint for Bert-48 and GPT-2 in different configurations. Since GPipe injects all the micro-batches at once, the high activation memory cost of it leads to \textit{OOM} (Out of Memory) in all the configurations. PipeDream has the second highest memory consumption because up to $D$ versions of weights have to be stashed. PipeDream-2BW reduces the stashed weights versions to 2. However, for language models, the first stage usually has more weights than other stages since it includes an extra embedding layer. Also, the pipeline schedule of PipeDream-2BW and DAPPLE determines that the activation memory consumption on the first worker is the highest. This double imbalance causes that the first worker commonly has the peak memory consumption for both PipeDream-2BW and DAPPLE. Since PipeDream-2BW stashes two versions of weights, it incurs \textit{OOM} as pipeline stages get coarser. In contrast, the schedule of bidirectional pipelines in Chimera determines that it has a more balanced memory consumption as shown in Figure~\ref{memCons}, which is consistent with the analysis in Talbe~\ref{tab:pipeline-schemes}. With the lowest activations memory cost occurring on the first (also the last) worker (see Figure~\ref{pipelinescompare}), the excessive weights of the first stage can be amortized in Chimera. Thus, although Chimera maintains two model replicas, it still has a little lower peak memory consumption than DAPPLE (the state-of-the-art synchronous approach which maintains one copy of the model) for four out of six configurations in Figure~\ref{memCons}. Although GEMS achieves the lowest (and also balanced) memory consumption among all approaches, this is at the cost of loss of parallelism. Overall, Chimera is on par with the state-of-the-art for the peak memory consumption, with a more balanced usage among the workers. These results are consistent with our analysis in Table~\ref{tab:pipeline-schemes}.

\subsection{Parallel Scalability}
\label{sec:parallelscala}
We first find the best configuration for each approach, and compare their best performance in the test of weak scaling.

\subsubsection{Performance Optimization Space for the Baselines} Given the mini-batch size $\hat{B}$ and the number of workers $P$, the best configuration of $B$ (micro-batch size), $D$ ( pipeline stages), and $W$ (the number of replicated pipelines) is not obvious a priori because of the trade-offs (i.e., computational efficiency and bubbles, \textit{allreduce} and \textit{p2p} communication overhead). We search the space of the parameters ($W$, $D$, and $B$ (for power-of-two)) to find the best performance for each baseline. The results for Bert-48 on 32 GPU nodes are presented in Figure~\ref{bertbaselinetuning}.

For synchronous baselines (such as GPipe and DAPPLE), the value of $B$ affects both computational efficiency and the bubble ratio. The planner of DAPPLE~\cite{fan2021dapple} gives an answer for how to select the configuration of $W$ and $D$ based on the profiling information, but it is not clear for how to select the best $B$. From Figure~\ref{bertbaselinetuning} we can see the highest throughput of both DAPPLE and GPipe (with \textit{activation recomputation}) is achieved by ($W$=8, $D$=4, $B$=4), under which they hit the sweet spot for the trade-off between \textit{p2p} communication overhead and \textit{allreduce} communication overhead by ($W$=8, $D$=4), and the sweet spot for the trade-off between bubble ratio and computational efficiency by $B$=4 (and $N$=16). GEMS prefers a large $B$ for high computational efficiency since a smaller $B$ does not help a lot to reduce the bubble ratio, and therefore its best performance is achieved by ($W$=8, $D$=4, $B$=32).  

Asynchronous baselines (PipeDream-2BW and PipeDream) always prefer the maximum $B$ fitting in the device memory, since there is no bubble problem for them. Note that PipeDream conducts gradient synchronization across $W$ pipelines after each backward pass on a micro-batch, thus its $\hat{B}$ is limited by the maximum $B$. Since the frequent gradient synchronization of PipeDream leads to high \textit{allreduce} overhead, its best performance is achieved with a deeper pipeline than others, namely by ($W$=4, $D$=8, $\hat{B}$=48). PipeDream-2BW scales to large $\hat{B}$ by accumulating the gradients for more than $D$ micro-batches (i.e., $N$>=$D$), and its best performance is achieved by ($W$=8, $D$=4, $B$=16) with activation recomputation.

\begin{figure}[h!]
\centering\includegraphics[width=0.88\linewidth]{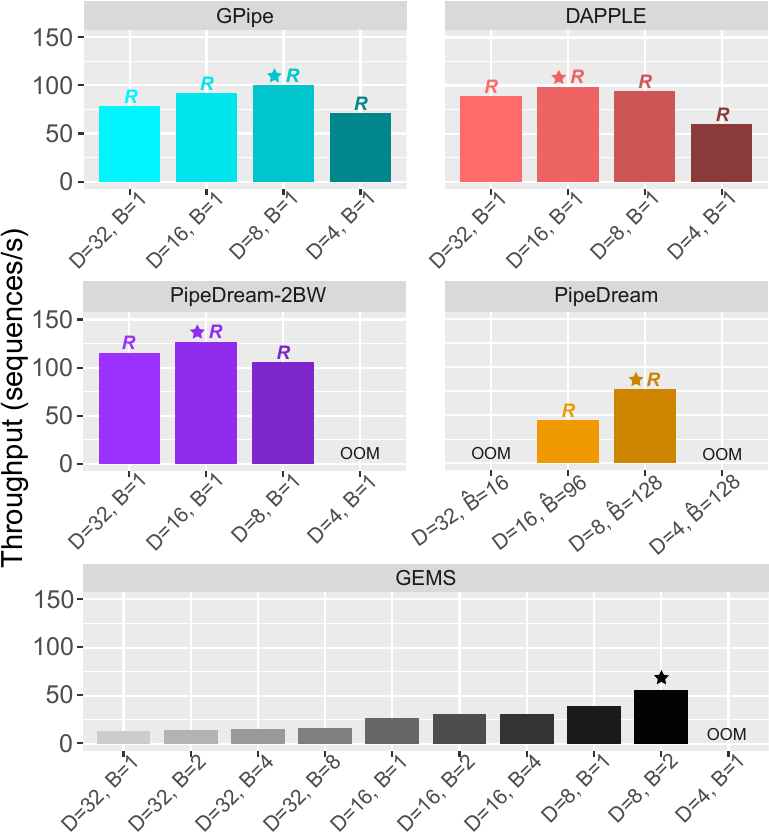}
\caption{\label{gptbaselinetuning} Performance tuning for the baselines for GPT-2 on 512 GPU nodes of Piz Daint. $\hat{B}$=512, except for PipeDream uses the maximum $\hat{B}$ fitting in the device memory. $R$ denotes activation recomputation to avoid \textit{OOM}. $Star$ marks the best performance.}
\end{figure}

For GPT-2, we present the performance tuning for each baseline by searching the parameter space in Figure~\ref{gptbaselinetuning}.

\begin{figure}[h!]
\centering\includegraphics[width=0.88\linewidth]{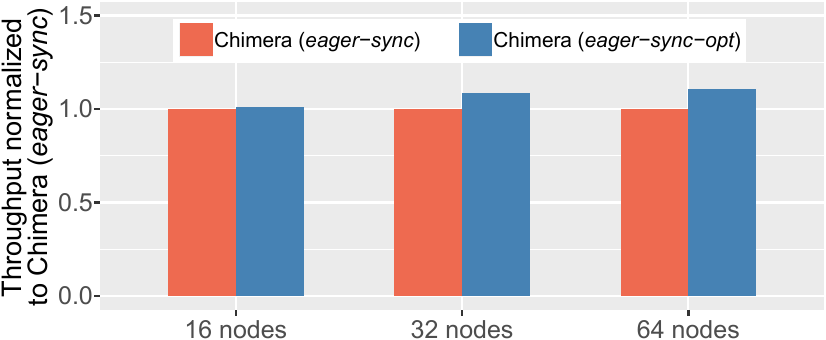}
\caption{\label{chimeracommopt} Throughput comparison between different gradient synchronization strategies of Chimera for Bert-48 on Piz Daint ($D$=4, $B$=8, and $\hat{B}$ scales from 256 to 1,024 as $P$ scales from 16 to 64).}
\end{figure}

\begin{figure}[h!]
\centering\includegraphics[width=0.88\linewidth]{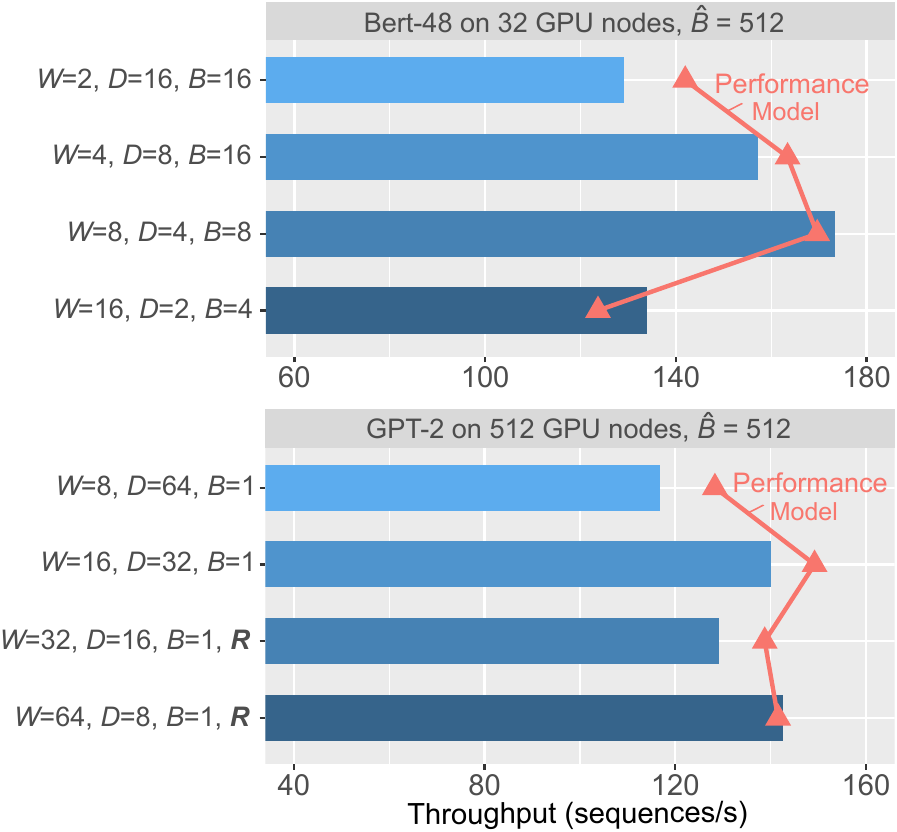}
\caption{\label{perfmodel} The practical and modelled throughput of Chimera on Piz Daint for Bert-48 on 32 GPU nodes with $\hat{B}$=256, and GPT-2 on 512 GPU nodes with $\hat{B}$=512.}
\end{figure}

\subsubsection{Performance Modelling of Chimera}
\label{sec:perfmodelexperiment}

We first evaluate the performance of Chimera with different gradient synchronization strategies discussed in Section~\ref{sec:commscheme}. We use \textit{eager-sync} to denote eager synchronization also conducted for the middle stages, and \textit{eager-sync-opt} to denote eager synchronization not conducted for the middle stages. Results in Figure~\ref{chimeracommopt} show that \textit{eager-sync-opt} achieves higher (e.g., 1.09x on 64 nodes) throughput than \textit{eager-sync}. These empirical results support our claim in Section~\ref{sec:commscheme}.

Figure~\ref{perfmodel} presents the practical training throughput of Chimera and the throughput predicted by the performance model (see Section~\ref{sec:perfmodel}). Note that since Chimera greatly alleviates the bubble problem, it greedily chooses the largest $B$ that fits in the device memory. The performance model is mainly used to select the best configuration of $W$ and $D$. Therefore, Chimera has a much smaller tuning space  compared with the synchronous baselines. The error of the performance model (see Equation~\ref{equ:perfmodel}) is within 10\% for both Bert-48 and GPT-2. For Bert-48, the performance model accurately selects the best configuration, i.e., $W$=8 and $D$=4. For GPT-2, the performance model selects $W$=16 and $D$=32, but the best performance is achieved by $W$=64 and $D$=8. However, the best performance is only 1.7\% higher than the one selected by the model. The inaccurate prediction is mainly because our model may overestimate the cost of activation recomputation used with $W$=64 and $D$=8. Although these two configurations achieve very close performance for GPT-2, it is worth mentioning that $D$=8 works better when scaling to large mini-batches because of less computation and \textit{p2p} communication overhead, while $D$=32 works better when scaling to more machines because of less gradient synchronization overhead. 

\subsubsection{Comparison with the Best Performance}
\label{sec:bestperf}

\begin{figure}[ht]
\centering\includegraphics[width=0.88\linewidth]{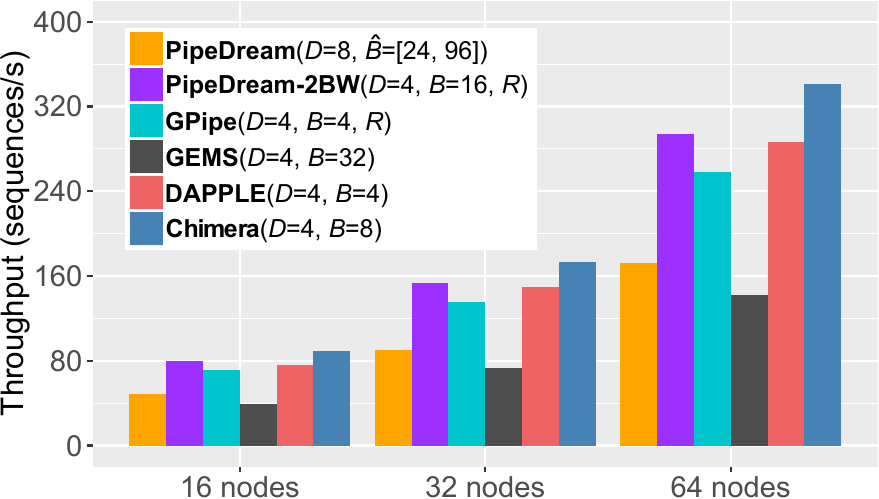}
\caption{\label{bertscalability} Weak scaling for Bert-48 on Piz Daint. As $P$ scales from 16 to 64, $\hat{B}$ scales from 256 to 1,024, except for PipeDream whose $\hat{B}$ scales from 24 to 96.}
\end{figure}

\begin{figure}[ht]
\centering\includegraphics[width=0.88\linewidth]{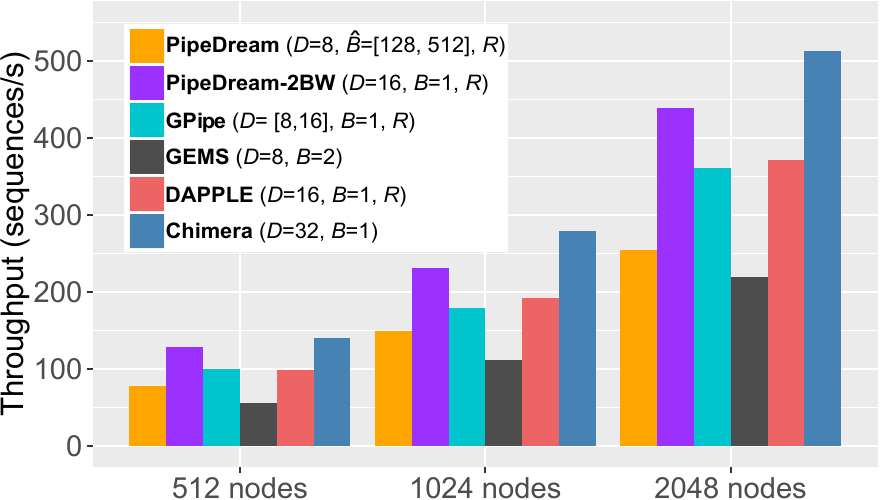}
\caption{\label{gptscalability} Weak scaling for GPT-2 on Piz Daint. As $P$ scales from 512 to 2,048, $\hat{B}$ scales from 512 to 2,048, except for PipeDream whose $\hat{B}$ scales from 128 to 512.}
\end{figure}

Figures~\ref{bertscalability} and ~\ref{gptscalability} present the results of weak scaling on Bert-48 and GPT-2, respectively. For all the baselines we present the best performance after searching the parameter space at different scales. Especially, to achieve the best performance, GPipe switches from $D$=8 to $D$=16 for GPT-2 on more than 512 GPU nodes. For Chimera, we present the practical throughput using the best configuration predicted by the performance model. The configuration used by each approach for the best performance is annotated in the legends of the figures.

For both Bert-48 and GPT-2, Chimera outperforms all the baselines at all scales. For Bert-48 on 64 nodes, Chimera outperforms PipeDream and PipeDream-2BW (asynchronous approaches) by 1.94x and 1.17x, respectively, and outperforms GPipe, GEMS, and DAPPLE (synchronous approaches) by 1.32x, 2.41x, and 1.19x, respectively. PipeDream frequently synchronizes the gradients after each backward pass, which compromises the 
training throughput. PipeDream-2BW uses $B$=16 with recomputation to achieve the 
best performance. Although PipeDream-2BW does not have bubble problem, it may not have enough computation to fully overlap the gradient synchronization overhead. GEMS has the highest bubble ratio and therefore has lower throughput than the others. To achieve the best performance, GPipe and DAPPLE use $B$=4 to reduce the bubble ratio but at the cost of lower computational efficiency. In contrast, Chimera has low bubble ratio while using $B$=8 for higher computational efficiency, and therefore outperforms GPipe and DAPPLE. 

For GPT-2 on 2,048 nodes, Chimera outperforms PipeDream and PipeDream-2BW (asynchronous approaches) by 2.01x and 1.16x, respectively, and outperforms GPipe, GEMS, and DAPPLE (synchronous approaches) by 1.42x, 2.34x, and 1.38x, respectively. There are two major advantages of Chimera: (1) Chimera has a low bubble ratio; (2) benefiting from a balanced memory consumption (as discussed in Section~\ref{sec:memcons}), Chimera with $D$=32 fits in the device memory without activation recomputation, while all other approaches except GEMS require recomputation. 
Chimera outperforms PipeDream-2BW mainly because no recomputation is required, and outperforms GPipe and DAPPLE because of both less bubbles and no recomputation. Using 512 nodes as the baseline, Chimera achieves 91.4\% parallel efficiency on 2,048 nodes in weak scaling for GPT-2, which demonstrates the efficiency of the communication scheme used in Chimera.

Note that we use the same model partition method as the default setting in PipeDream-2BW, namely evenly partitioning the basic layers among the workers. Other model partition methods trying to balance the weights among the workers may help to reduce the peak memory consumption of PipeDream-2BW, but this is outside the scope of this paper. Generally, Chimera is on-par with PipeDream-2BW (the latest asynchronous approach) in terms of training throughput, but more convergence-friendly since there is no stale weights in Chimera. 

\begin{figure}[ht]
\centering\includegraphics[width=0.88\linewidth]{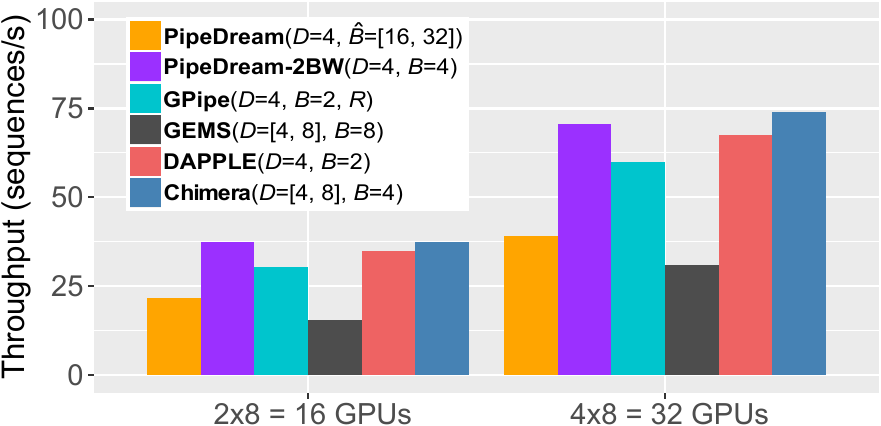}
\caption{\label{bertscalabilityv100} Weak scaling for Bert-48 on the cluster with 32 V100 GPUs. As $P$ scales from 16 to 32, $\hat{B}$ scales from 128 to 256, except for PipeDream whose $\hat{B}$ scales from 16 to 32. The max sequence length is set to 512.}
\end{figure}

We also conduct the evaluation on the cluster with 4x8=32 V100 GPUs connected by NVLink (intra-node) and Infiniband (inter-node). Experimental results for Bert-48 are shown in Figure~\ref{bertscalabilityv100}. On 32 V100 GPUs, Chimera improves the throughput by 1.10x-2.39x and 1.05x-1.89x over the synchronous and asynchronous pipeline approaches, respectively, which demonstrates that the same conclusions hold on newer machines.

\subsection{Scale to Large Mini-Batches on a Given Number of Machines}

In this section, we evaluate the training throughput when there are a large number of micro-batches available for each worker within a training iteration (i.e., $N>>D$), in which case the bubble problem of all synchronous approaches is alleviated.

\begin{figure}[ht!]
\centering\includegraphics[width=0.88\linewidth]{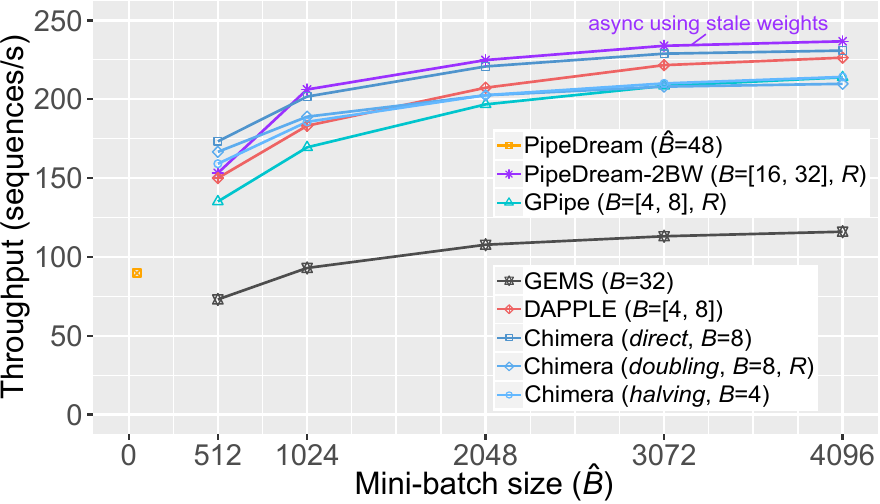}
\caption{\label{bertlargerbathces} Scale to large mini-batch size for Bert-48 on 32 GPU nodes of Piz Daint. $D$=8 for PipeDream, and $D$=4 for the others.}
\end{figure}

\begin{figure}[ht!]
\centering\includegraphics[width=0.88\linewidth]{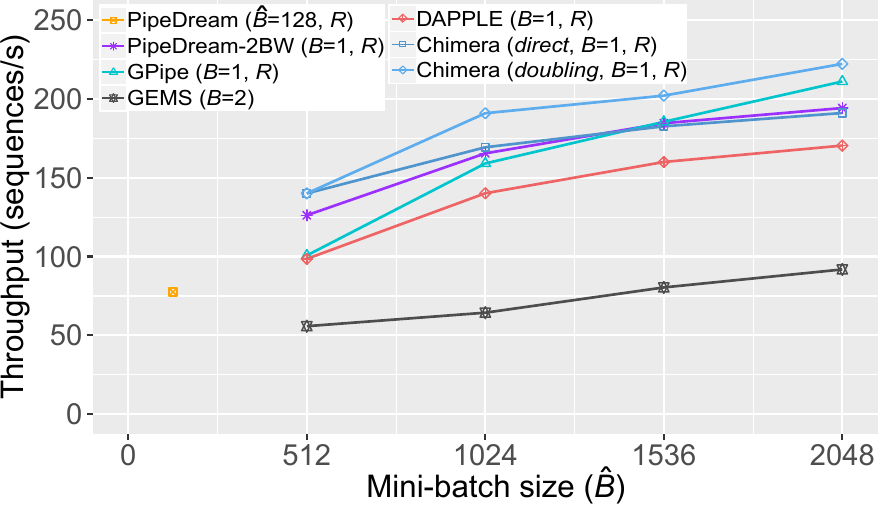}
\caption{\label{gptlargerbathces} Scale to large mini-batch size for GPT-2 on 512 GPU nodes of Piz Daint. DAPPLE and PipeDream-2BW switch from $D$=16 to $D$=8 when $\hat{B}$ >= 1,024, $D$=8 for the others.}
\end{figure}

Figure~\ref{bertlargerbathces} presents the throughput of Bert-48 on 32 GPU nodes when scaling to large mini-batches. Different from the other approaches, PipeDream updates the model after training on each micro-batch, and therefore its $\hat{B}$ stops scaling after reaching the memory limit. Consistently, we search the parameter space and present the best performance for each baseline. For example, to achieve the best performance, DAPPLE and GPipe switch from $B$=4 to $B$=8 when $\hat{B}$>=1,024 for higher computational efficiency, in which case the bubble problem is less important. PipeDream-2BW also uses $D$=4, and the $B$ increase from 16 to 32 when $\hat{B}>=1024$. 
Recall that in Section~\ref{sec:largerbatch} we discuss three methods for scaling Chimera to large mini-batches. Forward \textit{doubling} (with $B$=8) and backward \textit{halving} (with $B$=4) aim at solving the intermediate bubbles problem. However, the former suffers from recomputation overhead while the latter suffers from lower computational efficiency. \textit{Direct} concatenation (with $B$=8) achieves the best performance among these three methods on Bert-48, which can be explained by the fact that the intermediate bubbles caused by the uneven workloads between forward and backward passes can be utilized to accommodate the \textit{ptp} communication between pipeline stages. For $\hat{B}$<=2,048 where bubbles still matter, we observe significant improvement of Chimera (\textit{direct}) over all the synchronous approaches. Overall, for $\hat{B}$>=1,024, Chimera (\textit{direct}) is very close to PipeDream-2BW (asynchronous using stale weights), and achieves on average 1.13x, 2.07x, and 1.06x speedup over GPipe (suffering from recomputation), GEMS (suffering from high bubble ratio), and DAPPLE, respectively.

Figure~\ref{gptlargerbathces} presents the throughput of GPT-2 on 512 GPU nodes when scaling to large mini-batches. For GPT-2, Chimera ($D$=8) with forward \textit{doubling} outperforms \textit{direct} concatenation, since activation recomputation is required in both methods but the former removes intermediate bubbles. Note that  GPipe outperforms DAPPLE when scaling to large mini-batches in GPT-2, this is because both approaches require recomputation but the pipeline scheduling of GPipe is more regular and better to overlap the \textit{p2p} communication. 
Benefiting from the sophisticated (less bubbles and more communication overlap as discussed in Section~\ref{sec:largerbatch}) pipeline scheduling of Chimera with forward \textit{doubling}, our approach outperforms all the baselines, and achieves on average 1.13x, 1.18x, 2.60x, and 1.34x speedup over PipeDream-2BW, GPipe, GEMS, and DAPPLE, respectively. These results demonstrate that Chimera with forward \textit{doubling} efficiently scales to large mini-batches for the large models where activation recomputation is commonly required.

\subsection{Chimera with More than Two Pipelines}
\label{eval:morepips}
\begin{figure}[ht]
\centering\includegraphics[width=0.85\linewidth]{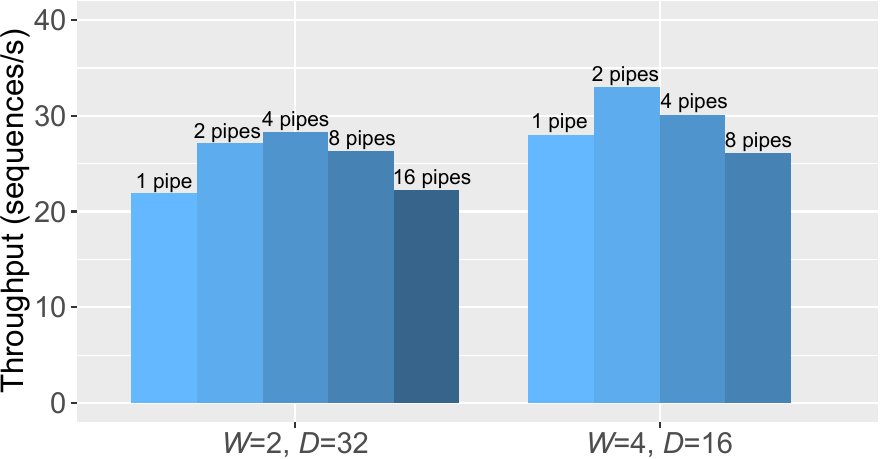}
\caption{\label{gptreplicas} The throughput of Chimera with more pipelines for a 32-layer GPT-2 with $\hat{B}$=64 on 64 GPU nodes of Piz Daint.}
\end{figure}

Figure~\ref{gptreplicas} presents the throughput of Chimera with more than two pipelines (i.e., model replicas) combined together on a 32-layer GPT-2 model. For the case of one pipeline, we use 1F1B scheduling with pipeline flushing. With $D$=32, four pipelines achieve the best performance because it hits the sweet spot of the trade-off between less bubbles and higher \textit{allreduce} overhead. However, as the pipeline stages get coarser by decreasing $D$ to 16, four pipelines performs worse then two pipelines because of the increasing of \textit{allreduce} overhead. Two pipelines (the default setting of Chimera) usually achieve the highest performance among all the configurations. We expect that Chimera with more than two pipelines would further improve the performance on future deep models with deeper pipeline and higher computation density on each stage.

\section{Conclusion}

Chimera brings new insights for efficiently pipelining large neural networks training at scale. Compared with the state-of-the-art pipeline approaches, Chimera achieves the best balance among pipeline efficiency, memory cost, and convergence friendliness. Empirical results for large language models training on up to 2,048 GPU nodes show that Chimera significantly improves the training throughput over the counterparts. We foresee that our approach will be one of the major solutions for massively scaling deep learning training. To reduce the communication cost of gradient synchronization by exploiting sparsification~\cite{hoefler2021sparsity,renggli2019sparcml} and quantization~\cite{alistarh2017qsgd} in deep learning training is our next step.

\begin{acks}
This project has received funding from the European Research Council (ERC) under the European Union's
Horizon 2020 programme (grant agreement DAPP, No. 678880, EPiGRAM-HS, No. 801039, and MAELSTROM, No. 955513). We also thank the Swiss National Supercomputing Center for providing the computing resources and excellent technical support.
\end{acks}



\bibliographystyle{ACM-Reference-Format}
\bibliography{mybib}

\newpage

\appendix
\section{Appendix: Artifact Description/Artifact Evaluation}

\subsection{SUMMARY OF THE EXPERIMENTS REPORTED}
We evaluated Chimera on the CSCS Piz Daint supercomputer. Each Cray XC50 compute node contains a 12-core Intel Xeon E5-2690 CPU with 64 GB RAM, and one NVIDIA Tesla P100 with 16 GB memory. The compute nodes are connected by Cray Aries interconnect in a Dragonfly topology. We used GLOO in PyTorch as the distributed backend. We utilized the GPU for acceleration in all the experiments, as described in the paper. The source code of Chimera is as follows:

Artifact name: \textcolor{blue}{Chimera}

Persistent ID: \textcolor{blue}{\url{https://github.com/Shigangli/Chimera}}

\subsection{BASELINE EXPERIMENTAL SETUP, AND MODIFICATIONS MADE FOR THE PAPER}

\textit{Relevant hardware details}: CSCS Piz Daint supercomputer. Each Cray XC50 compute node contains a 12-core Intel Xeon E5-2690 CPU and one NVIDIA Tesla P100 GPU. The filesystem is Lustre.

\textit{Operating systems and versions}: SUSE SLES 11.3

\textit{Compilers and versions}: gcc 9.3.0

\textit{Applications and versions}: Bert, GPT-2

\textit{Libraries and versions}: PyTorch 1.6

\textit{Key algorithms}: stochastic gradient descent

\textit{Input datasets and versions}: Wikipedia dataset, WikiText-2 dataset

\textit{URL to output from scripts that gathers execution environment information}.

\url{https://www.dropbox.com/s/md1jlcn3sm5bl9x/environment.txt?dl=0}


\end{document}